\documentclass[letterpaper]{dae-handout}
\graphicspath{{./}{./FIGURES/}{./figures/}{../Figures/}{../../Figures/}{./images/}{./graphics/}}

\newcommand{\trversion}{v0.1}
\newcommand{\trfilename}{Semantic-Arrow-Part-I.tex}
\newcommand{\trdate}{02026-FEB-27}
\newcommand{\trshorttitle}{The Semantic Arrow of Time, Part~I}

\usepackage[T1]{fontenc}
\usepackage[utf8]{inputenc}
\usepackage[bitstream-charter]{mathdesign}
\DeclareSymbolFont{operators}   {OT1}{cmr} {m}{n}
\DeclareSymbolFont{letters}     {OML}{cmm} {m}{it}
\DeclareSymbolFont{symbols}     {OMS}{cmsy}{m}{n}
\DeclareSymbolFont{largesymbols}{OMX}{cmex}{m}{n}
\SetSymbolFont{operators}{bold} {OT1}{cmr} {bx}{n}
\SetSymbolFont{letters}  {bold} {OML}{cmm} {b}{it}
\SetSymbolFont{symbols}  {bold} {OMS}{cmsy}{b}{n}
\SetMathAlphabet{\mathit} {normal}{OT1}{cmr}{m}{it}
\SetMathAlphabet{\mathbf} {normal}{OT1}{cmr}{bx}{n}
\SetMathAlphabet{\mathsf} {normal}{OT1}{cmss}{m}{n}
\SetMathAlphabet{\mathtt} {normal}{OT1}{cmtt}{m}{n}

\usepackage{amsmath}
\usepackage{amsthm}
\usepackage{booktabs}
\usepackage{enumitem}
\usepackage{fancyhdr}
\usepackage{titletoc}
\usepackage{etoc}
\usepackage{lastpage}
\usepackage{xcolor}
\usepackage{tikz}
\usepackage{eso-pic}
\usepackage{url}
\usepackage{morefloats}
\usepackage{placeins}
\usetikzlibrary{positioning,shapes.geometric}

\providecommand{\citenamefont}[1]{#1}
\setcitestyle{numbers,square}

\newtheorem{theorem}{Theorem}[section]

\newtheorem{proposition}[theorem]{Proposition}

\newtheorem{definition}[theorem]{Definition}

\newcommand{\fito}{\textsc{fito}}

\fancypagestyle{plain}{%
  \fancyhf{}%
  \fancyfoot[L]{\small\textsc{\trshorttitle}}%
  \fancyfoot[C]{\small\thepage\ of \pageref{LastPage}}%
  \ifarxiv\else
  \fancyfoot[R]{\scriptsize\texttt{\trfilename}}%
  \fi
  \ifarxiv\else
  \fancyfoot[L]{\raisebox{-0.8in}{\tiny\texttt{\trversion\ -- \trdate}}}%
  \fancyfoot[R]{\raisebox{-0.8in}{\tiny\texttt{\trfilename}}}%
  \fi
}
\title{The Semantic Arrow of Time \\Part~I: From Eddington to Ethernet}
\author[Paul Borrill]{Paul Borrill, D\AE D\AE LUS}
\date{02026-FEB-27}

\begin{document}
\setcounter{page}{1}
\maketitle
\thispagestyle{plain}

\daemargintoc

\begin{center}
\large\itshape Why Physics Got Time Right and Computing Got It Wrong
\end{center}
\vspace{1em}

\begin{abstract}
\noindent
This is the first of five papers comprising \emph{The Semantic Arrow of Time},
a series that traces a single argument from the philosophy of physics through
link-level networking to the corruption of files, email, and human memory.

The argument begins here, with a claim: \emph{computing's arrow of time is
semantic, not thermodynamic.}  The direction in which meaning is preserved or
destroyed across transactions is not a consequence of the second law of
thermodynamics but of design choices embedded in protocol architectures since
Shannon's 1948 channel model.  These choices encode what we call the
\emph{Forward-In-Time-Only} (\fito{}) assumption---the commitment that
causation is irreversible, acyclic, and globally monotonic.

We trace the intellectual lineage of this assumption from Eddington's 1927
coinage of ``the arrow of time,'' through the Boltzmann--Loschmidt debate,
to the contemporary philosophy of physics: Price's time-symmetric ontology,
Smolin's temporal naturalism, Rovelli's relational quantum mechanics, and
Roberts's analysis of time-reversal symmetry.  We show that fundamental
physics is time-symmetric at the microscopic level, that the thermodynamic
arrow emerges from boundary conditions rather than fundamental law, and that
recent experimental demonstrations of indefinite causal order establish that
nature admits correlations with no well-defined temporal ordering.

We then identify the category mistake (Ryle, 1949): computing inherited
Newton's absolute background time---via Shannon, Lamport, and the
impossibility theorems---and encoded it as a semantic primitive.  The
\fito{} assumption is not a law of nature but a design choice, and
recognizing this dissolves apparent constraints that have shaped forty
years of distributed systems theory.

The remaining papers in the series develop the constructive alternative:
Part~II formalizes the semantics of Open Atomic Ethernet links;
Part~III examines RDMA's ``completion fallacy'';
Part~IV traces semantic corruption through file synchronization, email,
and human memory;
Part~V presents the Leibniz Bridge as a unified framework.
\end{abstract}

\FloatBarrier
\section[Introduction]{Introduction: The Two Arrows}
\label{sec:intro}

In 1927, Arthur Eddington introduced a phrase that would organize a century
of physics and philosophy:
\begin{quote}
\small
Let us draw an arrow arbitrarily.  If as we follow the arrow we find more
and more of the random element in the state of the world, then the arrow is
pointing towards the future\,\ldots\  I shall use the phrase `time's arrow'
to express this one-way property of time which has no analogue in
space.~\citep{eddington1927}
\end{quote}
Eddington's arrow was thermodynamic: the direction of increasing entropy.
For nearly a century, it has served as the canonical explanation of why
time appears to flow in one direction---why we remember the past and not
the future, why eggs break but do not unbreak, why the universe expands
from a low-entropy initial state.%
\marginalia{Penrose's Weyl Curvature Hypothesis~\citep{penrose1979}
offers a gravitational counterpart: the initial singularity has vanishing
Weyl curvature, while the final singularity is Weyl-dominated.
Both explanations ground the arrow in boundary conditions, not in
time-asymmetric laws.}

This paper argues that there is a second arrow---one that has been hiding
in plain sight inside every distributed system, every network protocol,
and every file synchronization algorithm built since Shannon's 1948
paper~\citep{shannon1948}.  We call it the \emph{semantic arrow of time}:
the direction in which meaning is preserved or destroyed across
transactions.

The semantic arrow is not thermodynamic.  It does not track entropy.
It tracks whether a sequence of operations preserves the intended
interpretation of data---whether a file remains consistent after
synchronization, whether an email arrives with its causal context
intact, whether a distributed transaction commits atomically or
leaves torn state behind.  When the semantic arrow is violated,
the result is not heat death but \emph{semantic corruption}: data
structures that parse correctly but mean the wrong thing, timestamps
that impose an ordering no participant intended, memories that
reconstruct plausibly but inaccurately.

The thesis of this series is that computing's temporal architecture
embeds a specific, falsifiable assumption about the direction of
causation---the \fito{} assumption---and that this assumption is
a \emph{category mistake} in the sense of Ryle~\citep{ryle1949}:
it treats an epistemic convention (the logical ordering of messages)
as an ontic commitment (a law of physical causality).  The
consequence is that distributed systems theory has spent forty years
deriving impossibility theorems---FLP~\citep{flp1985}, Two
Generals~\citep{gray1978}, CAP~\citep{brewer2000,gilbert2002}---that
are theorems about \fito{} systems, not about physics.

\FloatBarrier
\section[Eddington's Legacy]{Eddington's Legacy: The Thermodynamic Arrow}
\label{sec:eddington}

The thermodynamic arrow of time rests on the second law: in an isolated
system, entropy does not decrease.  This statistical regularity, first
formalized by Clausius in 1865 and given a molecular-kinetic
interpretation by Boltzmann in 1872, provides the only widely accepted
physical basis for distinguishing past from future.%
\marginalia{Clausius: ``The entropy of the universe tends to a maximum.''
Boltzmann's $H$-theorem derived entropy increase from the kinetics of
ideal gases, but required the \emph{Stosszahlansatz} (molecular chaos
hypothesis)---itself a time-asymmetric assumption.}

Yet the microscopic laws of physics are time-symmetric.  Newton's
equations, Maxwell's equations, the Schr\"{o}dinger equation, and the
Einstein field equations are all invariant under time reversal
(or, more precisely, under CPT symmetry~\citep{roberts2022}).%
\marginalia{Roberts~\citep{roberts2022} provides the definitive
contemporary treatment of time-reversal symmetry in physics, showing
that the relevant symmetry is CPT (charge--parity--time) rather than
$T$ alone.}
The thermodynamic arrow therefore cannot derive from the fundamental
laws themselves; it must arise from something else.

The standard resolution, associated with Boltzmann and refined by
Penrose~\citep{penrose1979}, Callender~\citep{callender2017}, and
others, is that the arrow of time is a consequence of
\emph{boundary conditions}: the universe began in a state of
extraordinarily low entropy, and the second law is a statistical
consequence of evolving away from that improbable initial state.
The laws are symmetric; the arrow is emergent.%
\marginalia{Callender's \emph{What Makes Time Special?}~\citep{callender2017}
argues that time's distinctiveness is structural rather than
thermodynamic---a difference in how physical laws are ``tuned''
to distinguish temporal from spatial dimensions.}

This point---that the arrow is emergent, not fundamental---is the
first premise of our argument.  If the thermodynamic arrow derives
from initial conditions rather than from time-asymmetric laws, then
any system that \emph{assumes} an intrinsic temporal direction is
making an additional commitment beyond what physics requires.

\FloatBarrier
\section[Time-Symmetric Ontology]{The Case for Time-Symmetric Ontology}
\label{sec:price}

The most sustained philosophical argument for taking time-symmetry
seriously as an ontological commitment---not merely a formal property
of equations---has been developed by Huw Price over three
decades~\citep{price1996,price2012,price2025}.

Price's central move, articulated in \emph{Time's Arrow and
Archimedes' Point}~\citep{price1996}, is to adopt what he calls the
``Archimedean viewpoint'': a perspective that does not \emph{assume}
temporal asymmetry when examining the world's temporal structure.
From this standpoint, the default expectation is that the universe
should look the same whether we examine it in the forward or backward
temporal direction.  Any observed asymmetry requires explanation---it
cannot simply be presupposed.%
\marginalia{The Archimedean viewpoint is not a claim that the future
is determined or that we can see backward in time.  It is a
methodological principle: do not smuggle temporal asymmetry into
your explanatory framework before examining whether it is justified
by the phenomena.}

Price and Wharton~\citep{price-wharton2023} have recently extended
this argument to quantum mechanics, proposing that quantum
entanglement can be understood as a consequence of time-symmetric
causal structure---specifically, as a ``collider bias'' that emerges
when boundary conditions are imposed at both temporal endpoints of an
experiment.  If correct, this dissolves the apparent non-locality of
entanglement: what looks like instantaneous action at a distance is
actually a consequence of temporally symmetric constraints on the
mediating process.%
\marginalia{Price and Wharton's retrocausal model does not violate
signal locality.  No faster-than-light signaling is possible; the
correlations arise from the \emph{structure} of the boundary
conditions, not from superluminal influence.  See also the Stanford
Encyclopedia of Philosophy entry on retrocausality in quantum
mechanics~\citep{sep-retrocausality}.}

The significance for our argument is this: if the correct physical
ontology is time-symmetric, then any system that builds in a
preferred temporal direction is adding structure that nature does
not require.  As we shall see in Section~\ref{sec:leibniz}, this
is precisely the situation in which the Leibnizian principle of
the identity of indiscernibles applies: surplus ontological
structure signals a category mistake.

\FloatBarrier
\section[Temporal Naturalism]{Smolin's Temporal Naturalism}
\label{sec:smolin}

Lee Smolin approaches the arrow of time from the opposite
direction~\citep{smolin2013,smolin2013-tn}.  Where Price argues
for time-symmetry, Smolin argues that \emph{time is real}---that the
succession of present moments is the most fundamental feature of
reality, more fundamental than the laws of physics themselves.

Smolin's temporal naturalism makes three claims:
\begin{enumerate}[leftmargin=1.2cm]
  \item Time is real.  The passage from past to present to future is
    not an illusion or an artifact of coarse-graining.
  \item Laws of nature evolve.  They are not eternal Platonic objects
    but regularities that emerge from and change with the temporal
    structure of the universe.
  \item The universe is unique and causally closed.  There is no
    ensemble of universes; the laws must be explained by the history
    of the one universe that exists.
\end{enumerate}

At first glance, Smolin appears to oppose Price: one insists on
time-symmetry, the other on the reality of temporal passage.  But
the apparent conflict dissolves when we recognize what each is
denying.  Price denies that we should \emph{assume} temporal
asymmetry without justification.  Smolin denies that we should
\emph{eliminate} temporal passage as an illusion.  Both reject the
Newtonian/Einsteinian block universe in which past, present, and
future coexist timelessly.%
\marginalia{The block universe---the view that all events at all
times exist equally, and that temporal flow is an illusion---is the
standard interpretation of special relativity.  Both Price and
Smolin, from different directions, argue that this interpretation
imports metaphysical baggage beyond what the physics requires.}

For our purposes, Smolin's key contribution is the claim that laws
are not atemporal constraints imposed from outside but regularities
that acquire their meaning through their temporal context.  If
correct, this means that the \emph{semantic content} of a physical
law---what it means, what it permits, what it forbids---is
inseparable from the temporal structure in which it operates.  This
is the intuition that the semantic arrow captures at the level of
computing: the meaning of a transaction is inseparable from its
temporal structure.

\FloatBarrier
\section[Relational Time]{Rovelli's Relational Time}
\label{sec:rovelli}

Carlo Rovelli's relational quantum mechanics~\citep{rovelli1996}
and his thermal time hypothesis~\citep{connes-rovelli1994} offer a
third perspective.  For Rovelli, there are no observer-independent
facts: quantum states are relational, existing only with respect to
other systems.  The world is not made of substances but of
relations and interactions~\citep{rovelli2021}.

The thermal time hypothesis proposes that what we experience as
temporal flow is a consequence of our incomplete knowledge of the
microstate.  Given a thermodynamic state (a density matrix),
the ``flow of time'' is the one-parameter group generated by the
modular automorphism of the state~\citep{connes-rovelli1994}.
Different states generate different temporal flows; there is no
privileged ``true time'' independent of the observer's
thermodynamic situation.%
\marginalia{The thermal time hypothesis is technically precise:
it identifies the physical time parameter with the Tomita--Takesaki
modular flow of a KMS state.  But its philosophical import is
radical: time is not a background parameter but a consequence of
the relationship between an observer and the world.}

Rovelli's framework is a powerful ally of our argument in one respect:
it demonstrates that even the most sophisticated relational physics
can function without a preferred temporal direction.  But it also
contains a tension that Part~II of this series will address:
Rovelli treats time as \emph{emergent} from thermodynamic
coarse-graining, while the semantic arrow of time in computing is
\emph{constitutive}---it is the structure that makes transactions
meaningful.  If transaction timing is merely emergent, then there is
no principled basis for demanding atomicity; if it is constitutive,
then the design of link protocols must respect it as a primitive.

\FloatBarrier
\section[Indefinite Causal Order]{Indefinite Causal Order}
\label{sec:ico}

The most dramatic recent development in the physics of time is the
experimental demonstration of \emph{indefinite causal order}
(ICO)~\citep{oreshkov2012}.  Oreshkov, Costa, and
Brukner~\citep{oreshkov2012} showed in 2012 that quantum mechanics
permits correlations between laboratories that violate all causal
inequalities---correlations that cannot be explained by \emph{any}
fixed causal ordering of the laboratories' operations.

The ``quantum switch''---a device that places the order of two
quantum operations in superposition---has been realized
experimentally~\citep{rozema2024,vanderlugt2023}, and demonstrated
to provide computational advantages that are impossible with any
definite causal ordering~\citep{liu2024}.%
\marginalia{Rozema et al.~\citep{rozema2024} provide a comprehensive
review of experimental demonstrations of indefinite causal order.
Van der Lugt et al.~\citep{vanderlugt2023} achieved the first
device-independent certification of ICO in the quantum switch.}

Hardy's earlier work on quantum gravity
computers~\citep{hardy2007} had already established the theoretical
framework: if general relativity makes causal structure dynamical,
and if quantum mechanics makes dynamical variables indefinite, then
the causal structure of spacetime itself must admit superpositions.
Computation in such a regime cannot presuppose a fixed ordering of
operations---it must be defined relative to whatever causal structure
the physics permits.

For our argument, ICO establishes a decisive point: \emph{nature
does not require a globally consistent arrow of time.}  Any system
that assumes one---that encodes \fito{} as a primitive---is making
a commitment that goes beyond what physics permits.  This does not
mean that causal order is never well-defined; it means that the
assumption of global, definite causal order is an additional axiom,
not a consequence of known physics.

\FloatBarrier
\section[The Category Mistake]{The Category Mistake: From Physics to Computing}
\label{sec:category}

Gilbert Ryle introduced the concept of a \emph{category mistake}
to describe the error of representing facts of one logical type as
if they belonged to another~\citep{ryle1949}.  His famous example:
a visitor to Oxford is shown the colleges, the libraries, and the
playing fields, and then asks, ``But where is the University?''
The visitor has mistaken a collective abstraction for an entity of
the same logical type as its constituents.%
\marginalia{Ryle's category mistake has been applied more recently
to the hard problem of consciousness~\citep{category-electrified2023},
where treating ``experience'' as a substance rather than a process
is argued to be a categorical confusion.}

We now claim that distributed computing has committed a category
mistake of precisely this form.  The mistake is:

\begin{quote}
\emph{Treating the logical ordering of messages---an epistemic
convention for reasoning about system behavior---as if it were
a physical constraint on the causal structure of reality.}
\end{quote}

The genealogy of this mistake runs through three stages:

\paragraph{Stage 1: Shannon's Channel (1948).}
Shannon's mathematical theory of communication~\citep{shannon1948}
models information transfer as a unidirectional process: a sender
encodes a message, a channel corrupts it with noise, and a receiver
decodes it.  The model is \fito{} by construction---information
flows from sender to receiver, and the receiver's only recourse
against error is redundancy in the forward direction.
Shannon was modeling a specific engineering problem (telephone
transmission), and the unidirectional model was appropriate for
that purpose.  The category mistake occurs when the model is
\emph{promoted} from an engineering approximation to an ontological
claim about the structure of communication itself.%
\marginalia{Shannon himself was precise about the scope of his
theory: ``The fundamental problem of communication is that of
reproducing at one point either exactly or approximately a message
selected at another point.''  He did not claim that all
communication is unidirectional; the theory simply does not model
the reverse direction.}

\paragraph{Stage 2: Lamport's Happened-Before (1978).}
Lamport's happens-before relation~\citep{lamport1978} freed
distributed systems from dependence on synchronized clocks, but
it retained a stronger assumption: that the causal relationships
among events form a globally well-defined directed acyclic graph.
As shown in Part~I of the companion paper~\citep{borrill2026-lamport},
this encodes three implicit \fito{} commitments: temporal
monotonicity, asymmetric causation, and global causal reference.%
\marginalia{Lamport's third axiom---that the partial order
$\rightarrow$ is observer-independent---is stronger than what
special relativity permits.  Two observers in different inertial
frames do not agree on the temporal ordering of spacelike-separated
events.}

\paragraph{Stage 3: The Impossibility Theorems (1982--2002).}
FLP~\citep{flp1985} proved that no deterministic protocol can
guarantee consensus in an asynchronous system with even one faulty
process.  The Two Generals problem~\citep{gray1978} showed that no
finite number of messages can guarantee coordinated action across an
unreliable link.  CAP~\citep{brewer2000,gilbert2002} proved that a
distributed system cannot simultaneously guarantee consistency,
availability, and partition tolerance.

These results are universally presented as fundamental limits---laws
of distributed computing analogous to the laws of thermodynamics.
We argue that they are, instead, consequences of the \fito{}
assumption.  They are theorems about what is impossible
\emph{given} that causation is forward-only, acyclic, and globally
monotonic.  Drop the assumption, and the impossibility
disappears---not because the problems become easy, but because the
design space opens to include bilateral, reflective, and
interaction-derived causal structures that the \fito{} framework
cannot express.

\FloatBarrier
\section[The Leibnizian Principle]{The Leibnizian Principle Applied}
\label{sec:leibniz}

Leibniz's principle of the identity of indiscernibles states that
if two entities are empirically indistinguishable in all respects,
they are identical; any model that treats them as distinct contains
surplus structure.  Spekkens~\citep{spekkens2007} has applied this
principle to quantum mechanics via his toy model, showing that many
``quantum'' phenomena can be reproduced by a classical model
supplemented with an epistemic restriction (the knowledge balance
principle).  The lesson: features that appear ontological may be
epistemic artifacts of a model with too much structure.%
\marginalia{The knowledge balance principle: in any state of maximal
knowledge, the amount known about a system equals the amount that is
unknown.  This single constraint reproduces entanglement, no-cloning,
teleportation, and many other ``quantum'' phenomena in a classical
setting.  See Spekkens~\citep{spekkens2007}.}

The Leibnizian thread runs from Leibniz through Smolin (who calls
this the ``principle of no surplus structure'') through Spekkens
to the present argument.  Applied to distributed computing, it
yields:

\begin{proposition}[Surplus Structure in \fito{} Models]
\label{prop:surplus}
If two distributed system states are empirically indistinguishable
at every node---if no observation by any participant can distinguish
them---but the formal model assigns them different causal histories,
then the model contains surplus ontological structure.
\end{proposition}

This is precisely the situation with \fito{}-based models.  Two
events that are causally unrelated (neither can influence the other)
are nonetheless assigned an ordering by vector clocks or Lamport
timestamps.  The ordering is not observed; it is imputed by the
model.  The Leibnizian principle says: this imputed ordering is
surplus structure.  A model without it---one that admits indefinite
causal order between unrelated events---is more parsimonious and
more faithful to the physics.

\FloatBarrier
\section[The Semantic Arrow Defined]{The Semantic Arrow Defined}
\label{sec:semantic-arrow}

We are now in a position to define the central concept of this
series.

\begin{definition}[The Semantic Arrow of Time]
\label{def:semantic-arrow}
The \emph{semantic arrow of time} is the direction---within a
system of interacting processes---along which the intended
interpretation of shared state is preserved.  It is defined not by
entropy increase (thermodynamic arrow) or by subjective experience
(psychological arrow) but by the \emph{transactional structure}
of the system: the set of commitments, acknowledgments, and
reversibility conditions that determine whether a state transition
preserves or destroys meaning.
\end{definition}

The semantic arrow has the following properties:

\begin{enumerate}[leftmargin=1.2cm]
  \item \textbf{It is local, not global.}  Each transaction defines
    its own semantic arrow.  There is no requirement that all arrows
    align into a globally consistent direction.

  \item \textbf{It emerges from interaction, not from clocks.}
    The semantic arrow is determined by what the participants in a
    transaction \emph{do}---by the sequence of propose, reflect,
    commit, or abort---not by the timestamps attached to their
    messages.

  \item \textbf{It is reversible until commitment.}  Unlike the
    thermodynamic arrow, which is statistically irreversible, the
    semantic arrow can be reversed: a tentative state transition can
    be aborted, returning the system to its prior semantic state.
    Irreversibility enters only at the point of commitment.

  \item \textbf{Violation produces semantic corruption, not heat.}
    When the semantic arrow is violated---when a transaction
    half-commits, or when concurrent operations impose incompatible
    interpretations on shared state---the result is not a decrease
    in free energy but a loss of meaning: torn writes, phantom
    messages, silent data corruption, confabulated memories.
\end{enumerate}

The remaining papers in this series develop each of these properties
in detail.  Part~II shows how the Open Atomic Ethernet link state
machine~\citep{borrill2026-oae} realizes properties (1)--(3) at
the physical layer.  Part~III shows how RDMA's completion semantics
violate property~(2) at industrial scale.  Part~IV traces the
consequences of violated semantic arrows through file synchronization
(iCloud~\citep{borrill2026-icloud}), email, and human memory.
Part~V synthesizes the full framework as the Leibniz
Bridge~\citep{borrill2026-leibniz}.

\FloatBarrier
\section[The \fito{} Assumption in Computing]{The \fito{} Assumption in Computing}
\label{sec:fito-computing}

Before proceeding to the constructive parts of the series, we
catalogue the specific ways in which the \fito{} assumption manifests
in contemporary computing practice.  Each represents a point where
the category mistake---treating an epistemic convention as a physical
law---produces observable consequences.

\paragraph{Timeout-and-Retry (TAR).}
The universal failure-handling mechanism in distributed systems is
TAR: if a response does not arrive within a timeout period, retry
the request.  TAR assumes that the original request may have been
lost (forward failure) and that repeating it will eventually succeed
(forward recovery).  It cannot detect whether the original request
\emph{was} received and acted upon; the retry may therefore produce
duplicate effects.  The entire apparatus of idempotency tokens,
deduplication logs, and ``at-least-once'' vs.\ ``at-most-once''
delivery guarantees exists to patch the consequences of TAR's \fito{}
assumption.%
\marginalia{The once-delivery problem is analyzed in detail in
Borrill~\citep{borrill2026-wdcgw}.  TAR is ``the root of all evil''
in distributed systems: it converts every transient failure into a
potential semantic duplication.}

\paragraph{Clock Synchronization.}
NTP, PTP, Spanner's TrueTime, and every other clock synchronization
protocol assume that a shared notion of ``now'' can be established
by exchanging timestamped messages.  The \fito{} assumption enters
through the requirement that these messages be ordered by their
transmission times.  As Einstein showed, simultaneity is
frame-dependent; clock synchronization protocols paper over this
with bounded uncertainty intervals, but the underlying assumption
remains.%
\marginalia{Google's Spanner~\citep{corbett2013} uses GPS and
atomic clocks to bound uncertainty to $\sim$7\,ms.  This is
an engineering triumph, but it does not resolve the conceptual
problem: the protocol still assumes that ``before'' and ``after''
are well-defined across data centers separated by thousands of
kilometers.}

\paragraph{Last-Writer-Wins (LWW).}
When concurrent writes conflict, most cloud systems resolve the
conflict by keeping the write with the latest timestamp and
discarding the other.  LWW is \fito{} made explicit: the most
recent write (in forward time) is assumed to be the intended one.
But ``most recent'' is defined by a clock that may be
unsynchronized, and ``intended'' is a semantic property that no
timestamp can capture.  The result is silent data loss masquerading
as conflict resolution.%
\marginalia{iCloud's adoption of LWW and its consequences for file
synchronization are analyzed in Part~IV and in
Borrill~\citep{borrill2026-icloud}.  Parker~\citep{parker1983}
identified mutual inconsistency as a fundamental problem of
replicated data 40 years ago; LWW is the industry's response,
and it fails for exactly the reasons Parker predicted.}

\paragraph{Autoregressive Prediction.}
Large language models generate text by predicting the next token
given all preceding tokens---a process that is \fito{} by
construction.  The model has no mechanism to ``look back'' and
verify whether the tokens it has already emitted are consistent with
the tokens it is about to emit.  The result is hallucination:
locally fluent text that is globally incoherent, the semantic
analogue of a torn write in a distributed database.%
\marginalia{Part~IV develops the analogy between LLM hallucination
and semantic corruption in distributed systems.  Retrieval-augmented
generation (RAG) is a partial fix: it provides a ``backward''
verification step before generation.  But RAG still operates within
the autoregressive framework; it retrieves before generating, not
during.}

\FloatBarrier
\section[The Constructive Alternative]{Preview: The Constructive Alternative}
\label{sec:preview}

The remaining papers in this series develop an alternative to the
\fito{} framework, grounded in the physics and philosophy reviewed
above.  The core ideas, developed in detail in Parts~II--V, are:

\begin{enumerate}[leftmargin=1.2cm]
  \item \textbf{Interaction-derived causality.}  Causal order is not
    assumed \emph{a priori} but \emph{discovered} through
    interaction.  The direction of the semantic arrow emerges from
    the structure of the transaction, not from the timestamps on the
    messages.

  \item \textbf{Reflective acknowledgment.}  The minimum
    interaction required to establish causal order is a
    round-trip---a message and its reflection.  This is the
    network-level analogue of the Slowdown Theorem's requirement
    that measurement precision requires at least one
    round-trip~\citep{gorard-beggs-tucker}.

  \item \textbf{Reversibility until commitment.}  State
    transitions are tentative until both parties confirm.  The
    link state machine progresses through
    \textsc{tentative}~$\to$~\textsc{reflecting}~$\to$~\textsc{committed},
    with the option to abort at any point before commitment.

  \item \textbf{Indefinite Logical Timestamps.}  A four-valued
    causal structure that admits indefinite ordering between
    concurrent events, resolving only after symmetric
    link-level exchange~\citep{borrill2026-ilt}.

  \item \textbf{Mutual information conservation.}  The
    fundamental consistency primitive is not temporal precedence
    but the conservation of mutual information across a
    transaction---the Leibniz Bridge~\citep{borrill2026-leibniz}.
\end{enumerate}

\FloatBarrier
\section[Summary]{Summary}
\label{sec:summary}

We have argued that computing harbors a hidden arrow of time---not
the thermodynamic arrow of entropy increase, but a \emph{semantic}
arrow embedded in the \fito{} assumption that pervades every layer
of the networking and distributed systems stack.

The philosophical and physical evidence is clear:
\begin{itemize}[leftmargin=1.2cm]
  \item Fundamental physics is time-symmetric at the microscopic
    level (Roberts~\citep{roberts2022}).
  \item The thermodynamic arrow emerges from boundary conditions,
    not from time-asymmetric laws (Boltzmann, Penrose, Callender).
  \item Time-symmetric ontology is physically coherent and may
    explain quantum entanglement without non-locality
    (Price and Wharton~\citep{price-wharton2023}).
  \item The present moment has ontological status, and laws may
    evolve (Smolin~\citep{smolin2013}).
  \item Properties exist only at interaction; there is no
    observer-independent state of affairs
    (Rovelli~\citep{rovelli1996}).
  \item Nature admits correlations with no well-defined causal
    ordering (Oreshkov, Costa, and
    Brukner~\citep{oreshkov2012}).
\end{itemize}

Against this background, the \fito{} assumption in computing is
not a law of nature but a design choice---one that was appropriate
for Shannon's telephone channel but becomes a liability when
applied to distributed transactions, file synchronization, and
the preservation of meaning across interacting systems.

Recognizing the category mistake is the first step.  The remaining
four papers in this series develop the alternative.


\end{document}